\documentclass[letter,12pt]{article}
\usepackage{graphicx,amssymb,amsmath,epstopdf,color,hyperref}

\def\beq{\begin{equation}}
\def\eeq{\end{equation}}
\def\bea{\begin{eqnarray}}
\def\eea{\end{eqnarray}}

\def\gev{\, {\rm GeV}}

\newcommand{\gsim}{\lower.7ex\hbox{$\;\stackrel{\textstyle>}{\sim}\;$}}
\newcommand{\lsim}{\lower.7ex\hbox{$\;\stackrel{\textstyle<}{\sim}\;$}}

\def\xbibitem#1#2#3{\noindent\hangindent=2em #2\,(#3)}

\begin{document}

\begin{flushright}
DESY 16-176,
MCTP 16-20
\end{flushright}

\vspace{0.1in}

\noindent
\begin{center}
{\bf\Large The theoretical physics ecosystem \\ \vspace{0.1in}
behind the discovery of the Higgs boson} \\

\vspace{0.75cm}
{\large James D. Wells${}^{a,b,}$}\footnote{email: jwells@umich.edu}

{\it ${}^a$Deutsches Elektronen-Synchrotron DESY \\
Notkestra\ss e 85, D-22607 Hamburg, Germany}

{\it ${}^b$Michigan Center for Theoretical Physics (MCTP) \\
University of Michigan, Ann Arbor, MI 48109 USA}

(September 14, 2016)
\end{center}

\bigskip\bigskip
\noindent
{\it Abstract:} The discovery of the Higgs boson in 2012 was one of the most significant developments of science in the last half century. A simplified history has Peter Higgs positing it in the mid-1960s followed by a long wait while experimentalists progressively turned up collider energies until it appeared several decades later. However, in order for both the hypothesis and the experimental discovery to occur, a vast and complex theory ecosystem had to thrive in the years before Higgs's hypothesis and in the years that followed, building up to its discovery. 
It is further claimed that the Higgs boson hypothesis was an immoderate speculation, and therefore faith in theory argumentation and speculation was mandatory for the discovery program to proceed and reach its fulfillment. 
The Higgs boson could not have been discovered experimentally by accident. 

\vspace{0.7cm}
\begin{center}
{\it Lecture presented in part at the Kobayashi-Maskawa Institute for the Origin \\ of Particles and the Universe (KMI), University of Nagoya, Japan, 7 June 2016}
\end{center}

\vfill\eject
\tableofcontents
\bigskip

\section{Introduction}

A terse and deleteriously incomplete history of the Higgs boson says that it was postulated in 1964 by the theorist Peter Higgs~\cite{Higgs:1964pj} and then discovered in 2012 by experimentalists after a multi-decade herculean construction project at CERN to find it~\cite{Aad:2012tfa,Chatrchyan:2012xdj}. A narrative of the Higgs boson discovery that includes no discussion of theory work except for the hypothesis papers of Peter Higgs~\cite{Higgs:1964pj,Higgs:1964ia}, and maybe also of Brout and Englert's simultaneous work~\cite{Englert:1964et} on the subject, is a distortion of how science progresses, and what enables scientific discoveries.

The primary goal of this paper is, therefore, to elucidate the contributions of the theoretical physics community -- the ``theory ecosystem" -- to the Higgs boson hypothesis and to its later discovery. Of course, it goes without saying that experimental work was crucial, not just for the discovery data announced in 2012, but also through the knowledge attained in prior experiments and the skillful design of the discovery detectors Atlas and CMS. Nevertheless, the oft-repeated banal phrase ``Physics is an experimental science", which is only half correct since physics is equally a theoretical science,  indicates that there is no shortage of understanding of the importance of experiment to science in general, and to the discovery of the Higgs boson in particular. For this reason I focus on providing a somewhat comprehensive view of the theoretical physics contributions.

The discussion begins with an historical discussion of particle physics relevant for the context in which the Higgs boson hypothesis was formulated (sec.~\ref{sec:context}). After that a more focused discussion is given to theoretical physics efforts in the pre-hypothesis decades that gave rise directly to the Higgs boson hypothesis (sec.~\ref{sec:genesis}). The Higgs boson hypothesis was an immoderate speculation and was met with both acceptance and loathing by the community. The antipathy by which it was held is described in sec.~\ref{sec:speculation}, which is intended to impart to the reader  how uncertain and speculative the Higgs boson was viewed by many even up to the moment its discovery was announced. To discover the Higgs boson in experimental data required a tremendous amount of theoretical work, not just in making the hypothesis, but also after the hypothesis was initially articulated. The diverse and extensive theoretical physics ecosystem required for success of the entire endeavor is discussed in sec.~\ref{sec:ecosystem}. Conclusions are summarized in the final section.

\section{Context of the Higgs hypothesis}
\label{sec:context}

The known universe of visible matter, including our bodies, the earth, the sun, and everything we have ever seen in a laboratory is accounted for by the Standard Model (SM) of elementary particle physics. This theory says that there exist electrons, neutrinos, up-quarks and down-quarks as matter particles, which interact (i.e., experience forces) by the exchange of photons (electricity and magnetism), $W^{\pm}$ and $Z$ bosons (the ``weak interaction") and gluons (the ``strong interaction"). In addition to these particles there exists a second and a third family of matter particles that are exactly the same as the electrons, neutrinos, up-quarks and down-quarks in every way, except their masses are different. These particles include the charm, strange, top and bottom quarks, and the muon and tau leptons and their corresponding neutrinos.

The top quark was the last of these elementary particles to be discovered. Fermilab outside of Chicago took the honors of discovery in 1995~\cite{Abe:1995hr,Abachi:1994td}, and to this day it is the heaviest known elementary particle with mass of approximately $173\gev$. It is  near the mass of a Tungsten atom, which is not elementary and is made up of more  than 550 quarks bound together in its constituent protons and neutrons.

How did the top quark achieve such a high mass compared to, for example, the electron, which is more than 340,000 times lighter? How does the top quark  attain mass at all? For that matter, how do any of the elementary particles attain mass?  The answer that we know today is that a scalar boson exists -- the Higgs boson -- that has a background field value everywhere in space, and that other particles couple to this ``vacuum expectation value" (vev) of the field~\cite{Wells:2009kq}. 
The mass of a particle, such as the top quark or the electron, is directly proportional to its interaction strength with the vev. The top quark couples the strongest and therefore has the heaviest elementary particle mass (mass of top quark is $173\gev$), then the $Z$ boson (mass $91\gev$), then $W^\pm$ (mass $80\gev$), etc.\ down to the mass of the electron (mass $0.0005\gev$), and further down to the neutrinos below $10^{-10}\gev$. 

The question of how elementary particles get their mass had no good leads for quite some time even after the basics of forces and particles were surmised. For example, Glashow's 1961 study~\cite{Glashow:1961tr}, which is widely credited to be the first paper to articulate how the elementary particles and forces come together in a unified way, was cited in the 1979 Nobel Prize~\cite{Nobel:1979}  as the earliest paper of  the ``theory of the unified weak and electromagnetic interaction between elementary particles" (i.e., electroweak sector of the Standard Model). In that work the author did not have a good explanation for how masses come about. Instead he allowed the theory to break symmetries (i.e., retain only a ``partial symmetry").  For example, the author states that ``the part of the Lagrange function bilinear in the field variables which produces masses of the elementary particles need not be invariant under a partial-symmetry," and that ``the masses of [the gauge bosons] are as yet arbitrary"~\cite{Glashow:1961tr}. In other words, symmetry breaking masses are merely put in by hand.

It was not until the later work of Weinberg in 1967~\cite{Weinberg:1967tq} and Salam in 1968~\cite{Salam:1968rm} that connection was made between the theory of matter and forces with the spontaneous symmetry breaking insights of Higgs~\cite{Higgs:1964ia,Higgs:1964pj,Higgs:1966ev} and others~\cite{Englert:1964et,Guralnik:1964eu,Kibble:1967sv}.  Weinberg realized that a Higgs boson scalar field with a background vev could give mass to all the elementary particles.  In Weinberg's original paper he made this explicit by constructing a lagrangian that included all the matter particles ``plus a spin-zero doublet $\varphi=(\varphi^0~\varphi^-)$ [Higgs boson] whose vacuum expectation value will break $\vec T$ [$SU(2)_L$ gauge group generators] and $Y$ [hypercharge gauge group generator] and give the electron its mass"\cite{Weinberg:1967tq}.  

The Glashow-Weinberg-Salam (GWS) theory, as it came to be known, was now before the world to consider. Except for some important details, which also required deep theoretical insight, especially with regards to the strong interaction~\cite{Wilczek:1982yx,Wilczek:2002qt}, the structure of the SM was contained in these papers. In particular the hypothesis of a Higgs boson giving mass to all the elementary particles was clearly articulated. It was not immediately known if the hypothesis was correct. Indeed, it took over four decades to know that. And during that time there were many skeptics. Let us consider the challenges the scientific community had to this hypothesis and through this gain an understanding of how provocative the speculation was and how important the discovery of the Higgs boson has been in the history of science. But first, let us delve into the pre-Higgs world that set the groundwork for the hypothesis of the Higgs boson that was to come later.

\section{Genesis of the Higgs hypothesis}
\label{sec:genesis}

Landau's seminal 1937 paper~\cite{Landau:1937} should be considered the first identifiable pre-cursor theory to the Higgs boson.  Landau was in search of a way to characterize phase transitions in matter in a systematic way using thermodynamic potentials. He realized that the order parameter of a second-order phase transition -- the quantity that changes when a state goes from one phase to the next (e.g., total magnetization when transitioning to a ferromagnetic) -- is very small near the phase transition boundary. This calls out for a Taylor series expansion of the free energy near the transition point.  For example,  the ``Landau potential" can be written~\cite{Landau:1980} as
\beq
\label{eq:landau}
\Phi(P,T,\eta)=\Phi_0(P,T)+A(P,T)\eta^2+B(P,T)\eta^4
\eeq
where $P$ is the pressure, $T$ the temperature, $\eta$ the order parameter, and $A$ and $B$ are thermodynamics functions of pressure and temperature. For fixed pressure the Landau potential may take the approximate form
\beq
\Phi(T)=\frac{a}{2}(T-T_c)\,\eta^2+\frac{b}{4}\eta^4
\eeq
where $a$, $b$ and $T_c$ are positive constants. Minimizing this free energy leads to the solution
$\eta(T)=0$ when $T>T_c$ and $\eta(T)=a(T_c-T)/b$ when $T<T_c$. Thus, there is a critical temperature $T_c$ below which the phase transition has taken place.  At zero temperature the order parameter has the value $\eta_0=aT_c/b$.

The Landau free energy has been extremely useful in the history of physics, and played a central role in the development of many types of phase transitions witnessed. For example, the theory of superconductivity was elucidated by the application of Landau's theory of phase transitions to the superconducting state~\cite{Ginzburg:1950}.  The order parameter $\eta$ in that case is the number of superconducting charge carriers in the material.  In ferromagnetism the order parameter is the magnetization of the material.  At high temperatures the magnetic dipoles in a substance are not aligned and thermal fluctuations do not allow any non-zero values of the magnetization, but as the temperature drops, and thermal fluctuations become less destructive to the formation of a more ordered state, and the free energy of the substance is minimized in the ferromagnetic state.

Now, it was never considered to promote the order parameter to a new fundamental entity. It is considered a book-keeping device to take into account collective behavior of a system, which gives rise, through ordering, to some macroscopic phenomenon such as magnetization in a ferromagnet. Fluctuations of the order parameter throughout the system were studied, but those fluctuations were not fundamentally new objects but rather variations in the collective behavior of already known fundamental objects.

A good example of how the order parameter has been perceived in theories of phase transitions is in the theory of superconductivity.  Ginzburg and Landau in 1950~\cite{Ginzburg:1950}, using Landau's 1937 original theory of phase transitions~\cite{Landau:1937}, developed a complete mean-field theory of superconductivity using a simple Taylor series expansion of the order parameter (density of superconducting charge carriers) as in eq.~\ref{eq:landau}. This new approach led to many new successes, including the introduction of a coherence length, the understanding of quantization of magnetic fluxes and vortices, and a partial understanding of type I and type II superconductors based on the relation between coherence length and penetration depth (of the Meissner effect). However, again, nobody thought of the order parameter as a fundamental new state in nature with no previously known constituents.

The order parameter did turn out to represent, however, a new kind of dynamics in the system. Of course this was known to be the case. After all, what gives rise to superconductivity is a special kind of collective behavior, which for example is different than the behavior of freezing, and the Landau theory is agnostic to the precise nature of that behavior. But it is not odd or strange that in time, upon deeper probes into the system, the precise nature of the collective behavior would be resolved.  No new exotic fundamental particles were expected, but rather a new way of organizing the already well-known particles.

This is indeed what happened. The superconducting charge carriers of the Ginzburg-Landau theory were nothing more than a long-range pairing of electrons in the lattice. The BCS theory of superconductivity~\cite{BCS} elucidated the technical microscopic behavior and it did not require the introduction of exotic previously unknown new states -- only electrons and their normal electromagnetic interactions.

The theory of superconductivity, from both the Ginzburg-Landau perspective and the BCS perspective, was highly influential in the development of the Higgs hypothesis. It was noticed that the photon could achieve mass in these theories (Meissner effect) despite electricity and magnetism still being a good symmetry. Nambu~\cite{Nambu:1960tm,Nambu:1961tp} was arguably the first to understand this point~\cite{Close:2013}, as well as Anderson~\cite{Anderson:1963} (see also Vaks and Larkin~\cite{Vaks:1960}). This was recognized as spontaneous symmetry breaking, and it is synonymous with the ability to introduce a Landau free energy to describe the breaking of the symmetry (i.e., the order parameter taking on a non-zero value).

While work proceeded on  relativistic analogs to spontaneous symmetry breaking of the kind Landau pioneered, particle physicists were wrestling with other problems that put them on a path to make use of the symmetry breaking studies that were being carried out. Namely, particle physicists were finding matter and interactions that looked be described by theories similar to QED (i.e., relativistic gauge theory descriptions), but the analogs to the photon (the $W^\pm$ and $Z^0$ bosons) were massive. Naturally, therefore, one looked to the developments  of the research on spontaneous symmetry breaking for inspiration on how to give mass to force carriers but not break the symmetries that these force carriers represented. $W^\pm$ and $Z^0$ are force carriers for $SU(2)_L\times U(1)_Y$ and so a superconductivity analog was needed for breaking these symmetries. 

The Higgs hypothesis was born~\cite{Higgs:1964ia,Higgs:1964pj,Englert:1964et,Guralnik:1964eu,Kibble:1967sv} in the midst of all these particle physics struggles to give mass to vector bosons in relativistic gauge theories in analogy to superconductivity.  Indeed, the connection to superconductivity is made explicit by Higgs himself in his original 1964 paper, where he states that ``this phenomenon [Higgs mechanism of spontaneous symmetry breaking] is just the relativistic analog of the plasmon phenomenon to which Anderson has drawn attention"~\cite{Higgs:1964pj}, quoting Anderson's 1963 paper on the subject~\cite{Anderson:1963}.

To be explicit, the Higgs hypothesis, as we define it here, is the introduction of a quantum field, call  it the Higgs field $H(x)$, which serves as an order parameter for the spontaneous symmetry breaking from an $SU(2)\times U(1)_Y$ standard model phase into the $U(1)_{em}$ phase. Order parameters must be charged under the symmetries that they break, and so this Higgs field must be charged under $SU(2)_L\times U(1)_Y$. The vev of the Higgs field is the non-zero value it takes at low temperatures. 

When applied directly to the issues of explaining mass in particle physics, the Higgs hypothesis solved many problems. The ``partial symmetries" of Glashow~\cite{Glashow:1961tr} turned into full symmetries spontaneously broken. Furthermore the theory was proved to be renormalizable~\cite{'tHooft:1971rn}, giving it further credibility with the fashions of the time. And finally, a single field -- a single order parameter -- was all that was required to give masses to all elementary particles, from electrons to quarks to the $W^\pm,Z^0$ bosons. These were the tremendous successes of combining the ideas of spontaneous symmetry breaking with the problem of explaining elementary particle masses. However, none of these successes required the Higgs boson to be a fundamental particle. Indeed, such an idealized and ``naive" conception of the Higgs boson became the target of many attacks in the theory community.
In the next section we review these attacks that called into question the Higgs boson's existence as a stand-alone fundamental particle. 

\section{Higgs hypothesis as an immoderate speculation}
\label{sec:speculation}

It is important first to make a distinction between the Higgs mechanism and the Higgs boson.  The Higgs mechanism refers to the process of spontaneous symmetry breaking. It assumes the existence of an order parameter that is zero in the symmetric phase and non-zero in the non-symmetric phase, which is therefore charged under the full symmetry group. Agreement to the Higgs mechanism as a description of reduced symmetry of the ground state in no way commits one to the existence of a fundamental, propagating Higgs boson. One may agree that the symmetry breaking is well modeled by a mean field-theory type of description, as in Landau's theory of phase transition.  One may even introduce the field $H(x)$ as the order parameter, and could be tempted to call it a ``Landau field" just to make explicit that one is not requiring $H(x)$ to be a fundamental scalar field with no constitutes that gave rise to it.  Thus, the Higgs mechanism can be thought of as merely the introduction of a condensing charged field $H(x)$ that most efficiently represents the symmetry breaking status of the theory (i.e., the masses) but has nothing definitive to say about the dynamics of symmetry breaking (fundamental scalar versus fermion bilinear condensates, etc.).

The hypothesis that the Higgs field of the Higgs mechanism is fundamental, meaning that it has no constituents and is as ``real" and indivisible as an electron or a top quark, is the extra step that completes the Higgs hypothesis. It is this step that is firmly in the realm of immoderate speculation, and that many did not believe.  Whereas nearly all experts agreed with the Higgs mechanism, a large number thought the Higgs boson was simplistic thinking.

Before detailing the level of skepticism of the Higgs boson, we make a few remarks regarding the speculative nature of the Higgs mechanism itself. Now this speculation is not radical; however, strictly speaking it was not needed. One could introduce the $W$ and $Z$ bosons with mass, and give mass to all the fermions in the theory ``by hand". This would be a manifestation of what Glashow~\cite{Glashow:1961tr} originally called ``partial symmetry" that we discussed earlier, since the explicit masses would break explicitly the $SU(2)_L\times U(1)_Y$ symmetries down to $U(1)_{EM}$.  This did not worry Glashow, any more than explicitly breaking of flavor $SU(2)_L\times SU(2)_R$ in the strong interaction sector. 
The partial symmetry approach by Glashow was remarkably successful, and was recognized as the first paper to correctly unify the electromagnetic interactions leading to the correct predictions of massive gauge bosons and neutral currents via massive $Z^0$ exchange.

Glashow's approach of ``partial symmetry" gave way to ``spontaneous symmetry breaking" when it was understood that a Landau field $H(x)$ could describe all the partial symmetries of Glashow and also allow for the theory to be renormalizable~\cite{'tHooft:1971rn}. Renormalizability was keenly valued given the dramatic successes of the elegant, and renormalizable theory of QED. However, starting with Wilson's renormalization group work in the 1970s~\cite{Wilson:1973jj} the community began to understand that renormalizability is an unnecessary requirement for a valid low-scale theory~\cite{Weinberg:2016kyd,Wells:2012eff}, and so this extra incentive to agree to the Higgs mechanism is somewhat less attractive with deeper knowledge of effective theories today.

Now, we come to the full ``Higgs boson hypothesis" -- the proposition that the Higgs boson field is elementary and there exists an elementary propagating Higgs boson associated with fluctuations of the order parameter. On the surface it appears preposterous. There is no known symmetry breaking in nature that is described by Landau theory formalism whose Landau field -- the order parameter -- is not merely a simplified description of collective phenomena of other known particles and interactions of the system.  As discussed earlier, in ferromagnetism it is atomic dipoles aligning in an energetically favorable configuration. In superconductivity it is the condensing of two electrons together into Cooper pairs. These are but two examples. The Higgs hypothesis radically says that it is not a collection of already known states that collectively give rise to symmetry breaking, but rather the book-keeping Landau field is really a elementary, fundamental entity of its own -- the Higgs boson field with its accompanying elementary Higgs particle which propagates the modulus of the condensate.  

It is not surprising that such an hypothesis was met with wide skepticism.  A significant amount of effort in particle physics theory went into debunking the Higgs hypothesis. Many individuals who received even part training in the theory of phase transitions of condensed matter systems recoiled at the Higgs hypothesis. Incredulousness from the condensed matter community about a fundamental Higgs boson remains even today~\cite{Anderson:2015}. Promoting a book-keeping device (Landau's order parameter field) into something elementary and fundamental and ``real" appears too naive to many.  

One of the original arguments against the Higgs boson hypothesis was Susskind in 1979~\cite{Susskind:1978ms} who declared boldly in the abstract ``We argue that the existence of fundamental scalar fields [elementary Higgs boson] constitutes a serious flaw of the Weinberg-Salam theory [Standard Model of Particle Physics]." The main problem with the Higgs hypothesis that he had was the quantum quadratic sensitivities to much higher Planck scale, destabilizing the masses of all elementary particles. For a review and discussion of these issues, see recent discussions in~\cite{Giudice:2008bi,Wells:2013tta,Wells:2016luz}.

Skepticism about the Higgs boson hypothesis continued to be widespread among top particle theorist. The influential theorist Howard Georgi wrote an essay, ``Why I would be very sad if a Higgs boson were discovered" in 1997~\cite{Georgi:1997}. In this essay he bemoans the problems that many ideas of electroweak symmetry breaking have, and promotes the idea of a strongly coupled gauge theory being the origin of electroweak symmetry breaking. 
\begin{quote}
``The only place left to look for a way out of this swamp [the variety of electroweak symmetry breaking theories], it seems to me, is in strongly interacting chiral gauge theories. Many talented theorists have thought about this.... There are surely wonders hidden in the subject of strongly interacting chiral gauge theories. If we are forced to deal with them to deal by physics at the $SU(2)\times U(1)$ breaking scale, we may find them. If instead a Higgs is discovered and the physics at the $SU(2)\times U(1)$ breaking scale can be described by perturbation theory, we probably never will. This would be the real source of my sadness if a Higgs were discovered. It would mean that nature had missed a chance to teach us some new and interesting field theory. Personally, I don't think that she would be so malicious"\cite{Georgi:1997}.
\end{quote}
The result of these reflections was a substantial effort by the author and his collaborators~\cite{Collins:1999rz,Chivukula:1998wd} on top quark condensation theories of electroweak symmetry breaking that do not predict the simple light Higgs boson that was ultimately found at the LHC.

Another esteemed theoretical physics, Nobel Laureate Martinus Veltman, was even more direct in saying that he did not believe in the Higgs boson. In his widely attended lecture series at CERN in 1997~\cite{Veltman:1997} he made several unambiguous statements to this effect. For example, in his first lecture he said
\begin{quote} 
``I think that while our nature is doing something at the level [which] today phenomenologically can be well described by the Higgs system. I think by the time you get there [when LHC runs] you will find something else. How will it be, I haven't the foggiest notion....\\

``Of course I have no idea, except that I don't believe that the Higgs system as such, as it is advertised at this point as part of the Standard Model. I really don't believe that"~\cite{Veltman:1997}.
\end{quote}
and in his fourth lecture he said
\begin{quote} 
``And so here I am, and if I ask my own soul what I want, I say, I will believe in the Higgs sector such as generated by that [sigma model discussion] --- I believe in that from the point of view of the symmetries that it has, not for its dynamical content [Higgs boson]. This somehow I don't believe"~\cite{Veltman:1997}.
\end{quote}
Veltman was not alone in his sentiments. Nevertheless, they are remarkably candid words that signify the confidence that this renowned physicist had that the Higgs boson as we understood it could not be correct.  In other words, he believed the Higgs hypothesis to be false.

Let us be more specific in the stated reasons why particle physicists were skeptical of the Higgs boson hypothesis. One of the most ardent voices against the Higgs hypothesis was Lane~\cite{Lane:2002wv}, who in lectures wrote the following list of reasons why it is naive to believe in the elementary Higgs boson (i.e., Higgs hypothesis):
\begin{quote}
\begin{enumerate}
\item ``Elementary Higgs models provide no {\it dynamical} explanation for electroweak symmetry breaking.
\item Elementary Higgs models are {\it unnatural}, requiring fine tuning of parameters
to enormous precision. 
\item Elementary Higgs models with grand unification have a {\it hierarchy} problem
of widely different energy scales.
\item Elementary Higgs models are {\it trivial.}
\item Elementary Higgs models provide no insight to {\it flavor} physics"~\cite{Lane:2002wv}. 
\end{enumerate}
\end{quote}
Most theoretical physicists agreed with this list, often in its entirety but usually at least partially, only their suggested remedies differed. Similar statements can be found in thousands of papers since the Higgs hypothesis was formulated in the 1960's. 

So far we have explained and emphasized the first objection in Lane's list. Each of the other objections in the list have spawned
significant research effort also.
 It is not within the scope of this work to detail further the perceived deficiencies of the Higgs hypothesis in its myriad ways. If the reader wishes to learn more on this subject they are advised to consult~\cite{Wells:2009kq}. 

Our purpose here is rather to convince the reader that the Higgs hypothesis was widely challenged among the best and the brightest, and thus constituted an immoderate speculation in the history of particle physics.  For example, there have been over 1200 publications devoted to ``technicolor" and ``top condensate" theories, and ``Higgsless" theories," which are related to attempts to explain symmetry breaking in elementary particles analogously to that of superconductivity, thereby solving Lane's ``dynamical" problem. There are also many hundreds of papers on composite Higgs theories, which deny the existence of an elementary Higgs boson and say that it must be a composite of other particles.

To illustrate what motivated these authors to study alternate theories of symmetry breaking, the authors in one of the seminal early publications on Higgsless theories began their discussion by telling the readers why the Higgs boson should be doubted:
\begin{quote}
``The last unresolved mystery of the standard model (SM) of particle physics is the mechanism for electroweak symmetry breaking (EWSB). Within the SM it is assumed that a fundamental Higgs scalar is responsible for EWSB. This particle has not been observed yet, and its presence raises other fundamental issues like the hierarchy problem (that is how to avoid large quantum corrections to the mass of a light scalar)"\cite{Csaki:2003zu}. 
\end{quote}
This paper was published in arguably the top journal in the field of particle physics theory (Physical Review Letters) and has many hundreds of citations, illustrating the respect the community had for the motivations and ideas of the authors.  Papers on Higgsless theories were published right up to, and even slightly after 
the discovery of the Higgs boson, which was announced on July 4, 2012.

Other approaches to making the Higgs boson hypothesis more palatable were to agree to it being an elementary particle, but requiring that it be embedded into a larger theory with other states and dynamics that work collectively to obviate the problems described by Lane above. The primary example of this is supersymmetry. Supersymmetry assumes that a symmetry exists between fermions and boson such that if a fermion exists, a corresponding boson -- a superpartner -- must also exist. Thus, all fermions of the Standard Model, such as electrons, neutrinos, quarks, must have scalar boson superpartners. Susskind's problem with naturalness of scalars is rectified by the fermions and bosons working together to cancel the large quantum corrections that were feared to destabilize scalar bosons. The trouble with supersymmetry is that it needs to be ``broken" -- meaning, the superpartner scalars must have higher masses than the fermions since they otherwise would have already been seen by experiment. This might sound disastrous to the supersymmetry hypothesis, but such a breaking of the symmetry is not awkward from the theory perspective, and excellent ideas abound. For more discussion about the fundamentals of supersymmetry see~\cite{Martin:1997ns}.
Another approach is to assume the existence of large or warped extra dimensions, which through exotic spatial geometry rectify many of the problems stated by Lane. For more discussion about extra dimensional theories, see~\cite{ExtraDimensionsPDG}.  

There were many other theories and variants of theories postulated whose intention was 
to allow the Higgs boson to be an elementary particle by embedding it into a more complex exotic new framework. As with efforts to deny the fundamental nature of the Higgs boson altogether, these approaches gained many adherents in the community. Not all adherents of these theories are motivated primarily by making the Higgs boson palatable, but most viewed it as at least an important argument in their favor.

In conclusion, as has been illustrated in this section, the theory community has been divided into several categories of belief in the Higgs boson hypothesis:
\begin{enumerate}
\item There is no discernible Higgs boson (e.g., Higgsless theories, technicolor, etc.).
\item The Higgs boson exists but is not elementary (e.g, top condensate, composite Higgs theories).
\item The Higgs boson is elementary but embedded in an exotic theory to make it viable (e.g, supersymmetry, extra dimensions).
\item The Higgs boson is elementary and there is nothing more needed.
\end{enumerate}
The discovery of the Higgs boson eliminates the first category, which proves that the speculation of the Higgs boson hypothesis in at least the effective theory of the weak scale is a valid hypotheses.

The remaining three categories 2, 3 and 4 are still viable. However, categories 2 and 3 were generally thought by many to imply that experimentalists should have found new exotic ``beyond the Standard Model" phenomena revealing themselves at energy scales already probed by the LHC.  So far that has not happened. With the increases in energy and luminosity of the LHC, there are still prospects for validation of categories 2 and 3, but at this point the provisionally validity of those ideas is under stress for not having been manifest at the LHC.  Category 4 remains provisionally valid, since all the couplings that have been measured of the Higgs boson with other Standard Model particles have been in line with Standard Model expectations. However, these measurements are at the $\sim 10\%$ level~\cite{Khachatryan:2016vau}, which is not good enough to be declared ``precision Higgs measurements". In short, achieving further confidence of category 4, or discoveries of category 2 or 3, requires much more experimental effort.

Let us conclude this section by revisiting the claim of the Higgs boson hypothesis being an immoderate speculation. In terms of the categories of belief in the Higgs boson hypothesis, the first two categories, if experiment had established them, would not necessarily be radical. Indeed, they would be just another example in nature of symmetry breaking from collective phenomena of constituent particles. Categories 3 and 4, on the other hand, constitute an immoderate speculation, since the order parameter is declared to be a new elementary particle field with no other deriving identity. Higgs's original hypothesis best fits into category 4, and is therefore an immoderate speculation. Furthermore, one could argue that categories 3 and 4 are likely the hardest to establish experimentally. For example, supersymmetric theories are in category 3, and it is well known how difficult it is to find experimental evidence for many forms of supersymmetry even if superpartner masses are not much heavier than the Higgs boson mass~\cite{Fan:2011yu}. However, category 1 is already ruled out by data, and category 2 is generally described by strong interactions that perhaps require less from experiment to establish or rule out in time. Thus, we could find ourselves at the end of the LHC runs where we are fairly confident that nature is described by either category 3 or 4, where the immoderate speculation of an elementary Higgs boson is to all practical purposes the only surviving description of nature.

\section{Post-hypothesis phenomenology}

The Brout-Englert-Higgs (BEH) mechanism was a breakthrough yet also a rather narrow insight explaining how a vector boson in relativistic quantum field theory could achieve mass while at the same time preserving the gauge symmetry. As discussed in previous sections this insight was related to the insights of Nambu, Anderson and others who understood that the electromagnetic gauge symmetry is preserved in a superconductor despite the photon obtaining mass (Meissner effect). 

Higgs postulated further that a propagating Higgs boson should come along with the BEH mechanism, and henceforth the speculative scalar particle was called a ``Higgs boson". However, neither Higgs nor others had much of an idea at first of how this new boson could be seen by experiment.  Many years of research went into this question, summarized well by the widely read book {\it The Higgs Hunter's Guide}~\cite{HHG}. What made the discussion regarding the Higgs boson particularly difficult is that its mass was not known a priori. It depends on a free parameter $\lambda$ whose value could not be measured any other way except through discovery of the Higgs boson itself. Let us explain this part in some more detail.

The vector bosons of the Standard Model, the $W^\pm$ and $Z^0$ bosons, obtain mass by the vacuum expectation value (vev) of the Higgs field $\langle H\rangle = v/\sqrt{2}$, where $H$ is the Higgs field and $v$ is its vev, and the $\sqrt{2}$ is merely a convenient normalization factor.  The masses of the vector bosons are
\beq
M_W^2=\frac{1}{4}g^2v^2~~~~{\rm and}~~~~M_Z^2=\frac{1}{4}(g'^2+g^2)v^2
\eeq
where $g'$ and $g$ are the gauge couplings of the $U(1)_Y$ hypercharge force and $SU(2)_L$ weak force. These gauge couplings are measured well by the interactions among the fermions and have the values of $g\simeq 0.65$ and $g'\simeq 0.35$.  The gauge boson masses have been measured well too, and their values are $M_W\simeq 80\gev$ and $M_Z\simeq 91\gev$. Given the measurements of $g$, $g'$, $M_W$ and $M_Z$, we have several ways of determining what the vev is, and all give the same consistent result: $v\simeq 246\gev$. 

Knowing the vev of the Higgs boson is not enough to know its mass, which is determined by the equation
\beq
M_H^2=2\lambda v^2
\eeq
where $\lambda$ is an unknown and unmeasured coupling of the Higgs to itself. This unknown parameter self-coupling of the Higgs is necessary since, analogous to other particles, the Higgs ultimately must give mass to itself by virtue of how strongly it couples to the Higgs field~\cite{Wells:2009kq}. This uncertainty of the Higgs boson mass value was recognized very soon after Higgs formulated the hypothesis of its existence.

What makes the experimental analysis of the Higgs boson even more difficult than not knowing its mass, was that its signatures are radically different depending on its mass. For example, a Higgs boson mass of $600\gev$ would decay essentially every time to a pair of $W$ bosons or pair of $Z$ bosons, with a small extra fraction going to top quark pairs, and then essentially no other useful or accessible branching fraction into any other states.  If the Higgs boson mass were a few tens of GeV, then it would essentially only decay into bottom quarks with some extra fraction going to tau leptons and then essentially no other important branching fraction. From an experimental point of view there is a very significant difference between a $b$ quark final state and a $Z$ boson final state in the detectors.  One had to be prepared for all possibilities of the Higgs boson mass all the way from $M_H\sim$ electron mass to $M_H\sim$ TeV. 

The uncertainty in the Higgs mass, and all the induced uncertainties and varieties that went along with that, made the situation quite challenging for experiment. An interesting paper in 1975 devoted to the study of experimental implications of the Higgs boson nearly a decade after it was hypothesized stated, 
\begin{quote}
``We should perhaps finish with an apology and a caution. We apologize to experimentalists for having no idea what is the mass of the Higgs boson, unlike the case with charm and for not being sure of its couplings to other particles, except that they are probably all very small. For these reasons we do not want to encourage big experimental searches for the Higgs boson, but we do feel that people performing experiments vulnerable to the Higgs boson should know how it may turn up"~\cite{Ellis:1975ap}.
\end{quote}

The Atlas and CMS experiments had to prepare themselves for any possibility of the Higgs boson mass, including the richest and most complicated mass range of $120\gev\simeq M_H\simeq 160\gev$, which enabled competing branching fractions of the decays of the Higgs bosons into many states, such as $b$ quarks, $\tau$ leptons, $WW^*$ and $ZZ^*$. Note, $W^*$ means that the $W$ boson is ``off-shell" and we should think of it as the final state decay products of the $W$ (e.g., $\ell\nu$ and $\bar q'q$) rather than the $W$ itself. The meaning of $Z^*$ is same as for $W^*$ -- one should think of it only as its decay products.

This intermediate Higgs boson mass range of $120\gev\simeq M_H\simeq 160\gev$ required extra experimental care, since sharing its full decay branching fractions among multiple final states means that no one final state is necessarily the dominant signal of the Higgs boson. It also provides in time a variety of Higgs study opportunities, since the nature of the Higgs boson can be discerned better by having many accessible decay channels rather than just one or two.
Nevertheless, the initial Higgs boson discovery was ultimately made by two channels, $H\to \gamma\gamma$ and $H\to ZZ^*\to 4\ell$, where $4\ell$ means ``four leptons of either electron or muon variety" --- two of the leptons from $Z$ decays and the other two leptons from $Z^*$ decays. 

Recognizing that Higgs decays to two leptons could be a key discovery path required theory effort and insight from those trained best to discover it. Discussed in Ellis et al.~\cite{Ellis:1975ap} a decade after the Higgs hypothesis, it was solidified as a key search channel  in the 1988 study by Gunion et al.~\cite{Gunion:1987ke}, which parenthetically begins its abstract with the hedging phrase, ``If fundamental scalar bosons exist...". The paper opens with an historically illuminating comment that underscores our earlier claim here of how tentative the Higgs boson was viewed by the particle physics community even decades after its hypothesis:
\begin{quote}
``Although the Higgs mechanism is technically satisfactory to introduce gauge
boson and fermion masses in the standard model (SM), it is still not understood
physically, even two decades after it was introduced in particle physics. Some
theorists believe that no elementary scalar bosons (Higgs bosons) will exist.
Others, motivated by the belief that the standard model will be made conceptually
complete on the TeV scale by being part of a supersymmetric theory, expect
point-like scalars to be as much a part of the particle spectrum as fermions and
gauge bosons" ~\cite{Gunion:1987ke}.
\end{quote}
The authors here imply that there are two alternatives. One is that there are no naturally light fundamental scalars in nature, or there is supersymmetry which secures their existence through its fermion-boson symmetry. They sided with the latter and proceeded to study how it can be found, including through $h\to \gamma\gamma$ decays. One should note that these theorists were working out the phenomenology of a speculative idea, which in some addled quarters is considered an illegitimate or un-scientific exercise. Nevertheless, their work on the $\gamma\gamma$ signal was influential -- a signal that ultimately was required for the discovery to be made a quarter century later in 2012. 

In addition to theory work to determine what possible signals there might be to discovery the Higgs boson, there also was a requirement for detailed and rigorous calculations of the cross-sections and signal rates for these processes, and also the cross-sections for all background processes that mimicked the Higgs boson signal. 

For example, let us consider one of the most important discovery signals of the Higgs boson, which  was Higgs boson production via gluon fusion followed by decay of the Higgs boson to two photons, $pp\to gg\to H\to\gamma\gamma$.  The calculations that go into a full analysis of this process are difficult and the required techniques run broad and deep. The probability of finding two gluons at the right center of mass energy to make the Higgs boson required a community of theorists determining ``parton density functions", which were provided by several competing groups that had to relate hundreds of experimental results at colliders to their corresponding carefully computed theoretical values. This is notoriously subtle and it becomes a prodigious theoretical endeavor just to understand the errors that should be assigned to the density functions (see, e.g.,~\cite{Lai:2010nw}). 

In addition to the parton density functions, one needs also to compute the signal cross-sections. Computing $\sigma(gg\to H)$ is a one-loop process and rather straightforward, but computations at next order or next-to-next to leading order, etc., rapidly become career framing endeavors (see, e.g.,~\cite{Anastasiou:2012hx}).

Also, the background must be computed to very high accuracy. In our example case of $pp\to gg\to H\to \gamma\gamma$, the background comes mainly from a quark in one proton colliding with an anti-quark in another proton that then annihilate into two photons: $q\bar q\to \gamma\gamma$. The Higgs boson signal is then a ``bump" of two photons with invariant mass\footnote{The invariant mass of two photons is the square of their added four-momentum, which in this signal case should equal the Higgs boson mass, as required by Special Relativity.} equal to that of the Higgs boson, on top of a background of two photon events with invariant mass distribution continuously falling as a function of energy.  This continually falling background is required to be calculated at high-order in perturbation theory, requiring years of development and application of state-of-the-art techniques in quantum field theory (see, e.g.,~\cite{Bern:2001df}).

And finally, none of it is possible without taking these calculations and putting them into a computational framework that enables rapid and reliable calculations of parton density functions, signal processes and background processes.  There are a myriad of computer code packages that are developed by theoretical physicists that were required for the experimental analysis of data streaming into the experimental detectors Atlas and CMS.  These range from precision calculations of Higgs boson decays~\cite{Djouadi:1997yw} to energy distributions of background processes (e.g.,\cite{Mangano:2002ea}).

The theory work that goes into these different areas required a large community of theory scholars, each trained and working within their specialties. Only a small sample of the extensive nature of these studies has been provided in these section. In the next section additional evidence will be presented of the broad theory effort needed for the discovery of the Higgs boson. That evidence will come directly from the the two discovery papers written by Atlas~\cite{Aad:2012tfa} and CMS~\cite{Chatrchyan:2012xdj}.

\section{Theory and the Higgs boson discovery}
\label{sec:ecosystem}

In the previous section a discussion was presented on the development of the Higgs boson theory and phenomenology after the hypothesis of Higgs. What can be seen is that theoretical physics played a significant role in the discovery of the Higgs boson at many different levels and at many different times leading up to the discovery announcement of July 4, 2012.

In this section a brief analysis is made of the theoretical physics published papers that were cited by the Atlas or CMS discovery papers~\cite{Aad:2012tfa,Chatrchyan:2012xdj}. A complete list of these papers is given in the Appendix. Let us call these paper ``Higgs Discovery Theory References" (HDTRs). There are 115 of them altogether. In this section, these HDTR papers will be referred to by the last name of first author and its year, which can then be found in the appendix list of HDTRs. For example, the main hypothesis paper is Higgs (1964b).

By being included in the references of the experimental Higgs discovery paper, the HDTRs form a core set of papers that are widely recognized in the experimental community for directly influencing the efforts that lead to the Higgs boson discovery. Of course, there were many other articles and books that were deeply influential, which could have been acknowledged by the experimental groups. For example, the references of theory papers contained within all the HDTRs number in the thousands and were also crucial to the theory developments that led to the Higgs boson discovery.

The HDTRs can be subdivided into seven categories which are listed below. These categories represent distinctive areas of expertise in theoretical physics.
The number of papers belonging to each category is given in parentheses at the end of each category description.  

\begin{enumerate}
\item[C1.] Original hypothesis/speculation papers (7)
\item[C2.] SM foundations and the Higgs boson (7)
\item[C3.] Identifying Higgs boson signals (12)
\item[C4.] Higgs boson signal analysis (44)
\item[C5.] Background analysis (11)
\item[C6.] Collider physics theory (13)
\item[C7.] Computational tools (21)
\end{enumerate}

It must be kept in mind that each of these categories requires a community of scholars especially trained with the knowledge required to succeed in each. Some physicists are trained and have talent more for identifying theoretical and mathematical hypotheses (category C1), others for model building and theory fundamentals (category C2), others for translating subtle theory to experimental possibilities (C3), others for carrying out complex computations (C4 and C5), others for collider field theory (C6) and still others for computational integration of theoretical knowledge (C7). Some articles may straddle two or more categories, but each is put into only one category which is judged to be its most appropriate among the seven choices. The C category assigned to each paper is listed at the end of each of the HDTRs in the appendix.

The publication dates of the HDTRs are also interesting data to note. Here is a listing of number of articles in each stated time period:

\begin{enumerate}
\item[]1940-1949: 1
\item[]1950-1959: 1
\item[]1960-1969: 10
\item[]1970-1979: 8
\item[]1980-1989: 5
\item[]1990-1999: 13
\item[]2000-2004: 20
\item[]2005-2009: 30
\item[]2010-2012: 27
\end{enumerate}

The 14 C1 and C2 papers were in the 1960s and 1970s only, and they are the ones that dominate those early years. Important HDTR articles  in the C2 category where those by Glashow (1961), Salam (1968) and Weinberg (1967), which established the Standard Model of particle physics. The Weinberg article in particular is what put all the elements together and recognized the Higgs boson hypothesis as a final ingredient to give a self-consistent complete description of low-scale theory of elementary particle physics.

One further sees that a large fraction of HDTR papers were published within a few short years of the actual discovery, despite the hypothesis being nearly half a century old. The reason for this is that they represent the latest in the  long theoretical battle to calculate at increasingly better accuracy the signal and background processes required for discovery of the Higgs boson. There were many papers devoted to this in previous decades but they were continually supplanted (i.e., built upon) by more refined papers as the years went by and the Higgs boson had not shown up in one collider after another.  

Let us take for example Ferrera et al.\ (2011), which was released within a year of the Higgs boson discovery. This paper concerns itself with ``associated $WH$ production at hadron colliders" and presents a ``fully exclusive QCD calculation at NNLO."  The authors acknowledge that NNLO (next-to-next-to leading) QCD corrections are quite small, which may imply to some that they are not necessary to compute, especially given the tremendous effort required to come to reliable results. Nevertheless, as the authors point out, these corrections are important to know for the high-stakes and subtle Higgs boson search:
\begin{quote}
``The effect of NNLO QCD radiative corrections on the inclusive cross section is relatively
modest. However, it is important to study how QCD corrections impact the accepted cross
section and the relevant kinematical distributions. This is particularly true when severe selection
cuts are applied, as it typically happens in Higgs boson searches"~(Ferrara et al.\ 2011).
\end{quote}
The calculations involve ``a substantial amount of conceptual and technical complications" (Ferrara et al.\ 2011). This publication then cites numerous prior publications that did lower-order calculations, to which they can compare, and numerous publications that built upon the conceptual and technical knowledge over time on which they added, going back to 1991.

Two interesting points to note especially about the HDTR papers are, first, they are highly specialized and no small group of individuals could have ever developed all of them on their own, most especially the theorists who originated the hypothesis in the 1960s, and, second, every one of these articles was critical for the success of discovering the Higgs boson. The HDTR articles represent an extensive coverage of activities of theoretical particle physics, and yet they also are only the most proximate articles directly used by experimentalists in their search. The theory activity that enabled the subsequent production of the HDTRs is much greater, and, as we go back in time, one sees a vast succession of theoretical insights, work, and verification.

\section{Conclusions}
\label{sec:conclusions}

The Higgs boson discovery by the Atlas and CMS experiments in 2012 was an extraordinary achievement of particle physics. The achievement is equally impressive from the experimental and theoretical sides. On the experimental side, there is a long history of accelerator achievements and detector achievements that made the LHC accelerator and the Atlas and CMS detectors possible. Likewise, the experimental analysis methods and tools had to reach unprecedented sophistication in order to see the Higgs boson and have confidence in its discovery.

Equally impressive to the experimental efforts were the theory efforts. Theory progress did not start and end with the hypothesis of Higgs (and Brout and Englert and others) in the 1960s. They came about due to the extensive activity beforehand by theorists understanding phase transitions, in particular superconductivity, and by other theorists working out how vector bosons could have mass while retaining the gauge symmetries that they are associated with. Furthermore, the hypothesis would have been a dormant curiosity without the vast theoretical work that came after it.  It is astounding, and hard to understand by non-experts, that the hypothesized existence of a single particle coupled in a rather straightforward way to other particles, requires the work of thousands of theoretical physicists over decades to determine how it fits into the larger particle physics scheme of the Standard Model, to determine its various useful signatures in colliders, to determine how it might be mimicked by backgrounds, to develop computational tools, and to make calculations precise enough that a definitive discovery is made possible and realized.

The Higgs boson discovery took nearly fifty years to accomplish after its initial hypothesis. The time it takes to confirm an hypothesis of this kind is unconnected to the lifetime of humans, except indirectly due to the intense desire of those who saw the birth of the idea to confirm it, thus planning projects within their careers.  The time to discovery was ultimately dictated by the time it took to secure the financial resources and technical know-how, both experimentally and theoretically. This is the nature of all speculative hypotheses in science. There can be no dissuasive argument against an hypothesis based on how old the hypothesis is. The only effective arguments against speculative ideas are those that point to witnessed phenomena, or lack of phenomena, that are necessarily and unambiguously in direct contradiction to the idea. Otherwise, the speculative idea remains provisionally valid. The Higgs boson hypothesis remained provisionally valid for half a century, even though many researchers lost faith in it, and abandoned it altogether. In the end, the Higgs boson was there, and its valid status became secured.

Finally, the discovery of the Higgs boson is a triumph for theoretical physics, and a score for rationalism in its ongoing friendly with empiricism. Adding some prior experimental knowledge of nature to intensive pure thought created the Higgs boson hypothesis -- an impressive immoderate speculation that was ultimately confirmed. One heard from many over the years leading up to the turn-on of the LHC collider that the ``most exciting" prospect of the LHC would be that the Higgs boson is not there and that something we have never thought of before shows up.  For example, Wolfenstein \& Silva said, ``It will be even more exciting if the Higgs boson is not found, since that will mean there is still some fundamental new physics to be discovered"\cite{Wolfenstein:2010}. Similarly, Stephen Hawking said, ``But it [discovery of Higgs boson] is a pity in a way because the great advances in physics have come from experiments that gave results we didn't expect"~\cite{BBC:2012}.
These stated views have at their core a diminished view that theoretical physics reaches its height of usefulness and interest when it is merely an activity that reacts to and explains experimental surprises. However, theoretical physics is much more than that. As in the case of the Higgs boson we have discussed here, theoretical speculation and the theory ecosystem in which it is embedded make revolutionary discoveries possible to begin with, and that is where its extraordinary power resides.

\bigskip
\noindent
{\it Acknowledgments: } The author wishes to thank members of the Kobayashi-Maskawa Institute at Nagoya University for their kind invitation and stimulating hospitality during the visit when parts of this lecture were delivered. 

\vfill\eject
\section*{Appendix : Higgs Discovery Theory References}

This appendix contains all the references to peer-reviewed theoretical physics papers cited by the ATLAS~\cite{Aad:2012tfa} or CMS~\cite{Chatrchyan:2012xdj} Higgs boson discovery papers.

\bigskip

\xbibitem{Actis:2008ug}{Actis, S., G.~Passarino, C.~Sturm and S.~Uccirati,
  ``NLO Electroweak Corrections to Higgs Boson Production at Hadron Colliders,''
  Phys.\ Lett.\ B {\bf 670}, 12 (2008)
  doi:10.1016/ j.physletb.2008.10.018
  [arXiv:0809.1301 [hep-ph]].}{C4}

\xbibitem{Actis:2008ts}{Actis, S., G.~Passarino, C.~Sturm and S.~Uccirati,
  ``NNLO Computational Techniques: The Cases $H\to\gamma\gamma$ and $H\to gg$,"
   Nucl.\ Phys.\ B {\bf 811}, 182 (2009)
  doi:10.1016/j.nuclphysb.2008. 11.024
  [arXiv:0809.3667 [hep-ph]].}{C4}
  
\xbibitem{Aglietti:2004nj}{Aglietti, U., R.~Bonciani, G.~Degrassi and A.~Vicini,
  ``Two loop light fermion contribution to Higgs production and decays,''
  Phys.\ Lett.\ B {\bf 595}, 432 (2004)
  doi:10.1016/j.physletb.2004. 06.063
  [hep-ph/0404071].}{C4}

\xbibitem{Alioli:2008tz}{Alioli, S., P.~Nason, C.~Oleari and E.~Re,
  ``NLO Higgs boson production via gluon fusion matched with shower in POWHEG,''
  JHEP {\bf 0904}, 002 (2009)
  doi:10.1088/1126-6708/2009/04/002
  [arXiv:0812.0578 [hep-ph]].}{C4}
     
\xbibitem{Alwall:2007st}{Alwall, J.\ {\it et al.},
  ``MadGraph/MadEvent v4: The New Web Generation,''
  JHEP {\bf 0709}, 028 (2007)
  doi:10.1088/1126-6708/2007/09/028
  [arXiv:0706.2334 [hep-ph]].}{C7}
  
\xbibitem{Alwall:2011uj}{Alwall, J., M.~Herquet, F.~Maltoni, O.~Mattelaer and T.~Stelzer,
  ``MadGraph 5 : Going Beyond,''
  JHEP {\bf 1106}, 128 (2011)
  doi:10.1007/JHEP06(2011)128
  [arXiv:1106.0522 [hep-ph]].}{C7} 
  
\xbibitem{Anastasiou:2002yz}{Anastasiou, C.\ and K.~Melnikov,
  ``Higgs boson production at hadron colliders in NNLO QCD,''
  Nucl.\ Phys.\ B {\bf 646}, 220 (2002)
  doi:10.1016/S0550-3213(02)00837-4
  [hep-ph/0207004].}{C4}

\xbibitem{Anastasiou:2005qj}{Anastasiou, C., K.~Melnikov and F.~Petriello,
  ``Fully differential Higgs boson production and the di-photon signal through next-to-next-to-leading order,''
  Nucl.\ Phys.\ B {\bf 724}, 197 (2005)
  doi:10.1016/j.nuclphysb.2005.06.036
  [hep-ph/0501130].}{C4}
  
\xbibitem{Anastasiou:2008tj} {Anastasiou, C., R.~Boughezal and F.~Petriello,
  ``Mixed QCD-electroweak corrections to Higgs boson production in gluon fusion,''
  JHEP {\bf 0904}, 003 (2009)
  doi:10.1088/1126-6708/2009 /04/003
  [arXiv:0811.3458 [hep-ph]].}{C4}

\xbibitem{Anastasiou:2009kn}{Anastasiou, C., S.~Bucherer and Z.~Kunszt,
  ``HPro: A NLO Monte-Carlo for Higgs production via gluon fusion with finite heavy quark masses,''
  JHEP {\bf 0910}, 068 (2009)
  doi:10.1088/1126-6708/2009/10/068
  [arXiv:0907.2362 [hep-ph]].}{C7}
  
\xbibitem{Anastasiou:2012hx}{Anastasiou, C., S.~Buehler, F.~Herzog and A.~Lazopoulos,
  ``Inclusive Higgs boson cross-section for the LHC at 8 TeV,''
  JHEP {\bf 1204}, 004 (2012)
  doi:10.1007/JHEP04(2012)004
  [arXiv:1202.3638 [hep-ph]].}{C4}

\xbibitem{Arnold:2008rz}{Arnold, K.\ {\it et al.},
  ``VBFNLO: A Parton level Monte Carlo for processes with electroweak bosons,''
  Comput.\ Phys.\ Commun.\  {\bf 180}, 1661 (2009)
  doi:10.1016/j.cpc.2009.03.006
  [arXiv: 0811.4559 [hep-ph]].}{C7}
  
\xbibitem{Bagnaschi:2011tu}{Bagnaschi, E., G.~Degrassi, P.~Slavich and A.~Vicini,
  ``Higgs production via gluon fusion in the POWHEG approach in the SM and in the MSSM,''
  JHEP {\bf 1202}, 088 (2012)
  doi:10.1007/JHEP02(2012)088
  [arXiv:1111.2854 [hep-ph]].}{C4}

\xbibitem{Baglio:2010ae}{Baglio, J.\ and A.~Djouadi,
  ``Higgs production at the lHC,''
  JHEP {\bf 1103}, 055 (2011)
  doi:10.1007/ JHEP03(2011)055
  [arXiv:1012.0530 [hep-ph]].}{C4}
  
\xbibitem{Ball:2011mu}{Ball, R.D.\ {\it et al.},
  ``Impact of Heavy Quark Masses on Parton Distributions and LHC Phenomenology,''
  Nucl.\ Phys.\ B {\bf 849}, 296 (2011)
  doi:10.1016/j.nuclphysb.2011.03.021
  [arXiv: 1101.1300 [hep-ph]].}{C6}
  
\xbibitem{Barger:1990mn}{Barger, V.D., G.~Bhattacharya, T.~Han and B.~A.~Kniehl,
  ``Intermediate mass Higgs boson at hadron supercolliders,''
  Phys.\ Rev.\ D {\bf 43}, 779 (1991).
  doi:10.1103/PhysRevD.43.779.}{C4}
  
\xbibitem{Barr:2009mx}{Barr, A.J., B.~Gripaios and C.~G.~Lester,
  ``Measuring the Higgs boson mass in dileptonic $W$-boson decays at hadron colliders,''
  JHEP {\bf 0907}, 072 (2009)
  doi:10.1088/1126-6708/2009/ 07/072
  [arXiv:0902.4864 [hep-ph]].}{C4}
  
\xbibitem{Beenakker:2001rj}{Beenakker, W., S.~Dittmaier, M.~Kramer, B.~Plumper, M.~Spira and P.~M.~Zerwas,
  ``Higgs radiation off top quarks at the Tevatron and the LHC,''
  Phys.\ Rev.\ Lett.\  {\bf 87}, 201805 (2001)
  doi:10.1103/PhysRevLett.87.201805
  [hep-ph/0107081].}{C3}

\xbibitem{Beenakker:2002nc}{Beenakker, W., S.~Dittmaier, M.~Kramer, B.~Plumper, M.~Spira and P.~M.~Zerwas,
  ``NLO QCD corrections to $t\bar tH$ production in hadron collisions,''
  Nucl.\ Phys.\ B {\bf 653}, 151 (2003)
  doi:10.1016/S0550-3213(03)00044-0
  [hep-ph/0211352].}{C4}
  

\xbibitem{Binoth:2006mf}{Binoth, T., M.~Ciccolini, N.~Kauer and M.~Kramer,
  ``Gluon-induced W-boson pair production at the LHC,''
  JHEP {\bf 0612}, 046 (2006)
  doi:10.1088/1126-6708/2006/12/046
  [hep-ph/0611170].}{C5}
  
\xbibitem{Binoth:2008pr}{Binoth, T., N.~Kauer and P.~Mertsch,
  ``Gluon-induced QCD corrections to $pp\to ZZ\to l\bar ll'\bar l'$,"
  doi:10.3360/dis.2008.142
  arXiv:0807.0024 [hep-ph].}{C5}
  
\xbibitem{Bolzoni:2010xr}{Bolzoni, P., F.~Maltoni, S.~O.~Moch and M.~Zaro,
  ``Higgs production via vector-boson fusion at NNLO in QCD,''
  Phys.\ Rev.\ Lett.\  {\bf 105}, 011801 (2010)
  doi:10.1103/PhysRevLett.105.011801
  [arXiv:1003.4451 [hep-ph]].}{C4}

\xbibitem{Bozzi:2003jy}{Bozzi, G., S.~Catani, D.~de Florian and M.~Grazzini,
  Phys.\ Lett.\ B {\bf 564}, 65 (2003)
  doi:10.1016/ S0370-2693(03)00656-7
  [hep-ph/0302104].}{C4}
  
\xbibitem{Bozzi:2005wk}{Bozzi, G., S.~Catani, D.~de Florian and M.~Grazzini,
  ``Transverse-momentum resummation and the spectrum of the Higgs boson at the LHC,''
  Nucl.\ Phys.\ B {\bf 737}, 73 (2006)
  doi:10.1016/j.nuclphysb.2005.12.022
  [hep-ph/0508068].}{C4}
  
\xbibitem{Bredenstein:2006rh}{Bredenstein, A., A.~Denner, S.~Dittmaier and M.~M.~Weber,
  ``Precise predictions for the Higgs-boson decay $H\to WW/ZZ\to 4$\, leptons,"
  Phys.\ Rev.\ D {\bf 74}, 013004 (2006)
  doi:10.1103/ PhysRevD.74.013004
  [hep-ph/0604011].}{C4}
  
\xbibitem{Bredenstein:2006ha}{Bredenstein, A., A.~Denner, S.~Dittmaier and M.~M.~Weber,
  ``Radiative corrections to the semileptonic and hadronic Higgs-boson decays $H\to WW/ZZ\to 4$\, fermions,"
  JHEP {\bf 0702}, 080 (2007)
  doi:10.1088/1126-6708/2007/02/080
  [hep-ph/0611234].}{C4}

\xbibitem{Brein:2003wg}{Brein, O., A.~Djouadi and R.~Harlander,
  ``NNLO QCD corrections to the Higgs-strahlung processes at hadron colliders,''
  Phys.\ Lett.\ B {\bf 579}, 149 (2004)
  doi:10.1016/j.physletb.2003.10. 112
  [hep-ph/0307206].}{C4}
    
\xbibitem{Butterworth:1996zw}{Butterworth, J.M., J.~R.~Forshaw and M.~H.~Seymour,
  ``Multiparton interactions in photoproduction at HERA,''
  Z.\ Phys.\ C {\bf 72}, 637 (1996)
  doi:10.1007/s002880050286
  [hep-ph/9601371].}{C6}

\xbibitem{Cabibbo:1965zzb}{Cabibbo, N.\ and A.~Maksymowicz,
  ``Angular Correlations in $K_{e4}$ Decays and Determination of Low-Energy $\pi-\pi$ Phase Shifts,''
  Phys.\ Rev.\  {\bf 137}, B438 (1965)
  Erratum: [Phys.\ Rev.\  {\bf 168}, 1926 (1968)].
  doi:10.1103/PhysRev.137.B438, 10.1103/PhysRev.168.1926.}{C6}
  
\xbibitem{Cacciari:2007fd}{Cacciari, M.\ and G.~P.~Salam,
  ``Pileup subtraction using jet areas,''
  Phys.\ Lett.\ B {\bf 659}, 119 (2008)
  doi:10.1016/j.physletb.2007.09.077
  [arXiv:0707.1378 [hep-ph]].}{C6}
  
\xbibitem{Cacciari:2008gn}{Cacciari, M., G.~P.~Salam and G.~Soyez,
  ``The Catchment Area of Jets,''
  JHEP {\bf 0804}, 005 (2008)
  doi:10.1088/1126-6708/2008/04/005
  [arXiv:0802.1188 [hep-ph]].}{C6}
  
\xbibitem{Cacciari:2008gp}{Cacciari, M., G.~P.~Salam and G.~Soyez,
  ``The anti-$k_t$ jet clustering algorithm,''
  JHEP {\bf 0804}, 063 (2008)
  doi:10.1088/1126-6708/2008/04/063
  [arXiv:0802.1189 [hep-ph]].}{C6}

\xbibitem{Cacciari:2011ma}{Cacciari, M., G.~P.~Salam and G.~Soyez,
  ``FastJet User Manual,''
  Eur.\ Phys.\ J.\ C {\bf 72}, 1896 (2012)
  doi:10.1140/epjc/s10052-012-1896-2
  [arXiv:1111.6097 [hep-ph]].}{C7}

\xbibitem{Cahn:1983ip}{Cahn, R.N.\ and S.~Dawson,
  ``Production of Very Massive Higgs Bosons,''
  Phys.\ Lett.\ B {\bf 136}, 196 (1984)
  Erratum: [Phys.\ Lett.\ B {\bf 138}, 464 (1984)].
  doi:10.1016/0370-2693(84)91180-8.}{C3}

\xbibitem{Cahn:1986zv}{Cahn, R.N., S.~D.~Ellis, R.~Kleiss and W.~J.~Stirling,
  ``Transverse Momentum Signatures for Heavy Higgs Bosons,''
  Phys.\ Rev.\ D {\bf 35}, 1626 (1987).
  doi:10.1103/PhysRevD.35.1626.}{C4}
  
\xbibitem{Campbell:2006xx}{Campbell, J.M., R.~K.~Ellis and G.~Zanderighi,
  ``Next-to-Leading order Higgs + 2 jet production via gluon fusion,''
  JHEP {\bf 0610}, 028 (2006)
  doi:10.1088/1126-6708/2006/10/028
  [hep-ph/0608194].}{C4}

\xbibitem{Campbell:2011bn}{Campbell, J.M., R.~K.~Ellis and C.~Williams,
  ``Vector boson pair production at the LHC,''
  JHEP {\bf 1107}, 018 (2011)
  doi:10.1007/JHEP07(2011)018
  [arXiv:1105.0020 [hep-ph]].}{C5}

\xbibitem{Campbell:2011cu}{Campbell, J.M., R.~K.~Ellis and C.~Williams,
  ``Gluon-Gluon Contributions to $W^+ W^-$ Production and Higgs Interference Effects,''
  JHEP {\bf 1110}, 005 (2011)
  doi:10.1007/JHEP10(2011) 005
  [arXiv:1107.5569 [hep-ph]].}{C5}
  
\xbibitem{Catani:2003zt}{Catani, S., D.~de Florian, M.~Grazzini and P.~Nason,
  ``Soft gluon resummation for Higgs boson production at hadron colliders,''
  JHEP {\bf 0307}, 028 (2003)
  doi:10.1088/1126-6708/2003/07/ 028
  [hep-ph/0306211].}{C4}

\xbibitem{Choi:2002jk}{Choi, S.Y., D.~J.~Miller, M.~M.~Muhlleitner and P.~M.~Zerwas,
  ``Identifying the Higgs spin and parity in decays to Z pairs,''
  Phys.\ Lett.\ B {\bf 553}, 61 (2003)
  doi:10.1016/S0370-2693(02)03191-X
  [hep-ph/0210077].}{C4}
  
\xbibitem{Ciccolini:2003jy}{Ciccolini, M.L., S.~Dittmaier and M.~Kramer,
  ``Electroweak radiative corrections to associated WH and ZH production at hadron colliders,''
  Phys.\ Rev.\ D {\bf 68}, 073003 (2003)
  doi:10.1103/PhysRevD.68.073003
  [hep-ph/0306234].}{C5}

\xbibitem{Ciccolini:2007jr}{Ciccolini, M., A.~Denner and S.~Dittmaier,
  ``Strong and electroweak corrections to the production of Higgs + 2 jets via weak interactions at the LHC,''
  Phys.\ Rev.\ Lett.\  {\bf 99}, 161803 (2007)
  doi:10.1103/PhysRevLett.99.161803
  [arXiv:0707.0381 [hep-ph]].}{C4}
  
\xbibitem{Ciccolini:2007ec}{Ciccolini, M., A.~Denner and S.~Dittmaier,
  ``Electroweak and QCD corrections to Higgs production via vector-boson fusion at the LHC,''
  Phys.\ Rev.\ D {\bf 77}, 013002 (2008)
  doi:10.1103/ PhysRevD.77.013002
  [arXiv:0710.4749 [hep-ph]].}{C4}

\xbibitem{Corcella:2000bw}{Corcella, G., I.~G.~Knowles, G.~Marchesini, S.~Moretti, K.~Odagiri, P.~Richardson, M.~H.~Seymour and B.~R.~Webber,
  ``HERWIG 6: An Event generator for hadron emission reactions with interfering gluons (including supersymmetric processes),''
  JHEP {\bf 0101}, 010 (2001)
  doi:10.1088/1126-6708/2001/01/010
  [hep-ph/0011363].}{C7}
  
\xbibitem{Cornwall:1973tb}{Cornwall, J.M., D.~N.~Levin and G.~Tiktopoulos,
  ``Uniqueness of spontaneously broken gauge theories,''
  Phys.\ Rev.\ Lett.\  {\bf 30}, 1268 (1973)
  Erratum: [Phys.\ Rev.\ Lett.\  {\bf 31}, 572 (1973)].
  doi:10.1103/PhysRevLett.30.1268.}{C2}
  
\xbibitem{Cornwall:1974km}{Cornwall, J.M., D.~N.~Levin and G.~Tiktopoulos,
  ``Derivation of Gauge Invariance from High-Energy Unitarity Bounds on the S-Matrix,''
  Phys.\ Rev.\ D {\bf 10}, 1145 (1974)
  Erratum: [Phys.\ Rev.\ D {\bf 11}, 972 (1975)].
  doi:10.1103/PhysRevD.10.1145, 10.1103/PhysRevD.11.972.}{C2}
  
\xbibitem{Dawson:1990zj}{Dawson, S.,
  ``Radiative corrections to Higgs boson production,''
  Nucl.\ Phys.\ B {\bf 359}, 283 (1991).
  doi:10.1016/0550-3213(91)90061-2.}{C4}

\xbibitem{Dawson:2002tg}{Dawson, S., L.~H.~Orr, L.~Reina and D.~Wackeroth,
  ``Associated top quark Higgs boson production at the LHC,''
  Phys.\ Rev.\ D {\bf 67}, 071503 (2003)
  doi:10.1103/PhysRevD.67.071503
  [hep-ph/0211438].}{C4}
  
\xbibitem{Dawson:2003zu}{Dawson, S., C.~Jackson, L.~H.~Orr, L.~Reina and D.~Wackeroth,
  ``Associated Higgs production with top quarks at the large hadron collider: NLO QCD corrections,''
  Phys.\ Rev.\ D {\bf 68}, 034022 (2003)
  doi:10.1103/PhysRevD.68.034022
  [hep-ph/0305087].}{C4}
  
\xbibitem{deFlorian:2009hc}{de Florian, D.\ and M.~Grazzini,
  ``Higgs production through gluon fusion: Updated cross sections at the Tevatron and the LHC,''
  Phys.\ Lett.\ B {\bf 674}, 291 (2009)
  doi:10.1016/j.physletb. 2009.03.033
  [arXiv:0901.2427 [hep-ph]].}{C4}

\xbibitem{deFlorian:2011xf}{de Florian, D.\ G.~Ferrera, M.~Grazzini and D.~Tommasini,
  ``Transverse-momentum resummation: Higgs boson production at the Tevatron and the LHC,''
  JHEP {\bf 1111}, 064 (2011)
  doi:10.1007/JHEP11(2011)064
  [arXiv:1109.2109 [hep-ph]].}{C4}

\xbibitem{deFlorian:2012yg}{de Florian, D.\ and M.~Grazzini,
  ``Higgs production at the LHC: updated cross sections at $\sqrt{s}=8$ TeV,''
  Phys.\ Lett.\ B {\bf 718}, 117 (2012)
  doi:10.1016/j.physletb.2012.10.019
  [arXiv:1206.4133 [hep-ph]].}{C4}

\xbibitem{Degrassi:2004mx}{Degrassi, G.\ and F.~Maltoni,
  ``Two-loop electroweak corrections to Higgs production at hadron colliders,''
  Phys.\ Lett.\ B {\bf 600}, 255 (2004)
  doi:10.1016/j.physletb.2004.09.008
  [hep-ph/0407249].}{C4}
  
\xbibitem{Denner:2011mq}{Denner, A., S.~Heinemeyer, I.~Puljak, D.~Rebuzzi and M.~Spira,
  ``Standard Model Higgs-Boson Branching Ratios with Uncertainties,''
  Eur.\ Phys.\ J.\ C {\bf 71}, 1753 (2011)
  doi:10.1140/epjc/ s10052-011-1753-8
  [arXiv:1107.5909 [hep-ph]].}{C4}
  
\xbibitem{Denner:2011id}{Denner, A., S.~Dittmaier, S.~Kallweit and A.~Muck,
  ``Electroweak corrections to Higgs-strahlung off W/Z bosons at the Tevatron and the LHC with HAWK,''
  JHEP {\bf 1203}, 075 (2012)
  doi:10.1007/JHEP03(2012)075
  [arXiv:1112.5142 [hep-ph]].}{C4}
  
\xbibitem{Denner:2011rn}{Denner, A., S.~Dittmaier, S.~Kallweit and A.~Muck,
  ``EW corrections to Higgs strahlung at the Tevatron and the LHC with HAWK,''
  PoS EPS {\bf -HEP2011}, 235 (2011)
  [arXiv:1112.5258 [hep-ph]].}{C7}
  
\xbibitem{DeRujula:2010ys}{De Rujula, A., J.~Lykken, M.~Pierini, C.~Rogan and M.~Spiropulu,
  ``Higgs look-alikes at the LHC,''
  Phys.\ Rev.\ D {\bf 82}, 013003 (2010)
  doi:10.1103/PhysRevD.82.013003
  [arXiv:1001.5300 [hep-ph]].}{C4}
  
\xbibitem{Dittmar:1996ss}{Dittmar, M.\ and H.~K.~Dreiner,
  ``How to find a Higgs boson with a mass between 155-GeV - 180-GeV at the LHC,''
  Phys.\ Rev.\ D {\bf 55}, 167 (1997)
  doi:10.1103/PhysRevD.55.167
  [hep-ph/9608317].}{C3}
  
\xbibitem{Dixon:2003yb}{Dixon, L.J.\ and M.~S.~Siu,
  ``Resonance continuum interference in the diphoton Higgs signal at the LHC,''
  Phys.\ Rev.\ Lett.\  {\bf 90}, 252001 (2003)
  doi:10.1103/PhysRevLett.90.252001
  [hep-ph/0302233].}{C4}

\xbibitem{Djouadi:1991tka}{Djouadi, A., M.~Spira and P.~M.~Zerwas,
  ``Production of Higgs bosons in proton colliders: QCD corrections,''
  Phys.\ Lett.\ B {\bf 264}, 440 (1991).
  doi:10.1016/0370-2693(91)90375-Z.}{C4}

\xbibitem{Djouadi:1997yw}{Djouadi, A., J.~Kalinowski and M.~Spira,
  ``HDECAY: A Program for Higgs boson decays in the standard model and its supersymmetric extension,''
  Comput.\ Phys.\ Commun.\  {\bf 108}, 56 (1998)
  doi:10.1016/S0010-4655(97)00123-9
  [hep-ph/9704448].}{C7}

\xbibitem{Ellis:1975ap}{Ellis, J.R., M.~K.~Gaillard and D.~V.~Nanopoulos,
  ``A Phenomenological Profile of the Higgs Boson,''
  Nucl.\ Phys.\ B {\bf 106}, 292 (1976).
  doi:10.1016/0550-3213(76)90382-5.}{C3}
  
\xbibitem{Ellis:1987xu}{Ellis, R.K., I.~Hinchliffe, M.~Soldate and J.~J.~van der Bij,
  ``Higgs Decay to $\tau^+\tau^-$: A Possible Signature of Intermediate Mass Higgs Bosons at the SSC,''
  Nucl.\ Phys.\ B {\bf 297}, 221 (1988).
  doi:10.1016/0550-3213(88)90019-3.}{C3}

\xbibitem{Englert:1964et}{Englert, F.\ and R.~Brout,
  ``Broken Symmetry and the Mass of Gauge Vector Mesons,''
  Phys.\ Rev.\ Lett.\  {\bf 13}, 321 (1964).
  doi:10.1103/PhysRevLett.13.321.}{C1}

\xbibitem{Ferrera:2011bk}{Ferrera, G., M.~Grazzini and F.~Tramontano,
  ``Associated WH production at hadron colliders: a fully exclusive QCD calculation at NNLO,''
  Phys.\ Rev.\ Lett.\  {\bf 107}, 152003 (2011)
  doi:10.1103/PhysRevLett.107.152003
  [arXiv:1107.1164 [hep-ph]].}{C5}
  
\xbibitem{Figy:2003nv}{Figy, T., C.~Oleari and D.~Zeppenfeld,
  ``Next-to-leading order jet distributions for Higgs boson production via weak boson fusion,''
  Phys.\ Rev.\ D {\bf 68}, 073005 (2003)
  doi:10.1103/PhysRevD. 68.073005
  [hep-ph/0306109].}{C4}
  
\xbibitem{Frixione:2002ik}{Frixione, S.\ and B.~R.~Webber,
  ``Matching NLO QCD computations and parton shower simulations,''
  JHEP {\bf 0206}, 029 (2002)
  doi:10.1088/1126-6708/2002/06/029
  [hep-ph/0204244].}{C6}
  
\xbibitem{Frixione:2003ei}{Frixione, S., P.~Nason and B.~R.~Webber,
  ``Matching NLO QCD and parton showers in heavy flavor production,''
  JHEP {\bf 0308}, 007 (2003)
  doi:10.1088/1126-6708/2003/08/007
  [hep-ph/0305252].}{C6}
  
\xbibitem{Frixione:2005vw}{Frixione, S., E.~Laenen, P.~Motylinski and B.~R.~Webber,
  ``Single-top production in MC@NLO,''
  JHEP {\bf 0603}, 092 (2006)
  doi:10.1088/1126-6708/2006/03/092
  [hep-ph/0512250].}{C5}
  
\xbibitem{Frixione:2008yi}{Frixione, S., E.~Laenen, P.~Motylinski, B.~R.~Webber and C.~D.~White,
  ``Single-top hadroproduction in association with a W boson,''
  JHEP {\bf 0807}, 029 (2008)
  doi:10.1088/1126-6708/2008/07/029
  [arXiv:0805.3067 [hep-ph]].}{C5}

\xbibitem{Frixione:2010ra}{Frixione, S., F.~Stoeckli, P.~Torrielli and B.~R.~Webber,
  ``NLO QCD corrections in Herwig++ with MC@NLO,''
  JHEP {\bf 1101}, 053 (2011)
  doi:10.1007/JHEP01(2011)053
  [arXiv:1010.0568 [hep-ph]].}{C7}

\xbibitem{Gallicchio:2010sw}{Gallicchio, J.\ and M.~D.~Schwartz,
  ``Seeing in Color: Jet Superstructure,''
  Phys.\ Rev.\ Lett.\  {\bf 105}, 022001 (2010)
  doi:10.1103/PhysRevLett.105.022001
  [arXiv:1001.5027 [hep-ph]].}{C6}
  
\xbibitem{Gao:2010qx}{Gao, Y., A.~V.~Gritsan, Z.~Guo, K.~Melnikov, M.~Schulze and N.~V.~Tran,
  ``Spin determination of single-produced resonances at hadron colliders,''
  Phys.\ Rev.\ D {\bf 81}, 075022 (2010)
  doi:10.1103/PhysRevD.81.075022
  [arXiv:1001.3396 [hep-ph]].}{C6}
  
\xbibitem{Georgi:1977gs}{Georgi, H.M., S.~L.~Glashow, M.~E.~Machacek and D.~V.~Nanopoulos,
  ``Higgs Bosons from Two Gluon Annihilation in Proton Proton Collisions,''
  Phys.\ Rev.\ Lett.\  {\bf 40}, 692 (1978).
  doi:10.1103/PhysRevLett.40.692.}{C3}
  
\xbibitem{Glashow:1961tr}{Glashow, S.L.,
  ``Partial Symmetries of Weak Interactions,''
  Nucl.\ Phys.\  {\bf 22}, 579 (1961).
  doi:10.1016/0029-5582(61)90469-2.}{C2}
  
\xbibitem{Glashow:1978ab}{Glashow, S.L., D.~V.~Nanopoulos and A.~Yildiz,
  ``Associated Production of Higgs Bosons and Z Particles,''
  Phys.\ Rev.\ D {\bf 18}, 1724 (1978).
  doi:10.1103/PhysRevD.18.1724.}{C3}
  
\xbibitem{Gleisberg:2008ta}{Gleisberg, T., S.~Hoeche, F.~Krauss, M.~Schonherr, S.~Schumann, F.~Siegert and J.~Winter,
  ``Event generation with SHERPA 1.1,''
  JHEP {\bf 0902}, 007 (2009)
  doi:10.1088/1126-6708/2009/02/007
  [arXiv:0811.4622 [hep-ph]].}{C7}

\xbibitem{Golonka:2005pn}{Golonka, P.\ and Z.~Was,
  ``PHOTOS Monte Carlo: A Precision tool for QED corrections in $Z$ and $W$ decays,''
  Eur.\ Phys.\ J.\ C {\bf 45}, 97 (2006)
  doi:10.1140/epjc/s2005-02396-4
  [hep-ph/0506026].}{C7}
  
\xbibitem{Gray:2011us}{Gray, R.C., C.~Kilic, M.~Park, S.~Somalwar and S.~Thomas,
  ``Backgrounds To Higgs Boson Searches from $W \gamma^* -> l \nu l (l)$ Asymmetric Internal Conversion,''
  arXiv:1110.1368 [hep-ph].}{C5}

\xbibitem{Gunion:1987ke}{Gunion, J.F., G.~L.~Kane and J.~Wudka,
  ``Search Techniques for Charged and Neutral Intermediate Mass Higgs Bosons,''
  Nucl.\ Phys.\ B {\bf 299}, 231 (1988).
  doi:10.1016/0550-3213(88)90284-2.}{C3}
    
\xbibitem{Guralnik:1964eu}{Guralnik, G.S., C.~R.~Hagen and T.~W.~B.~Kibble,
  ``Global Conservation Laws and Massless Particles,''
  Phys.\ Rev.\ Lett.\  {\bf 13}, 585 (1964).
  doi:10.1103/PhysRevLett.13.585.}{C1}

\xbibitem{Hamberg:1990np}{Hamberg, R., W.~L.~van Neerven and T.~Matsuura,
  ``A Complete calculation of the order $\alpha-s^{2}$ correction to the Drell-Yan $K$ factor,''
  Nucl.\ Phys.\ B {\bf 359}, 343 (1991)
  Erratum: [Nucl.\ Phys.\ B {\bf 644}, 403 (2002)].
  doi:10.1016/0550-3213(91)90064-5.}{C6}
  
\xbibitem{Han:1991ia}{Han, T.\ and S.~Willenbrock,
  ``QCD correction to the $pp\to WH$ and $ZH$  total cross-sections,''
  Phys.\ Lett.\ B {\bf 273}, 167 (1991).
  doi:10.1016/0370-2693(91)90572-8.}{C5}

\xbibitem{Harlander:2002wh}{Harlander, R.V.\ and W.~B.~Kilgore,
  ``Next-to-next-to-leading order Higgs production at hadron colliders,''
  Phys.\ Rev.\ Lett.\  {\bf 88}, 201801 (2002)
  doi:10.1103/PhysRevLett.88.201801
  [hep-ph/0201206].}{C4}

\xbibitem{Higgs:1964ia}{Higgs, P.W.,
  ``Broken symmetries, massless particles and gauge fields,''
  Phys.\ Lett.\  {\bf 12}, 132 (1964a).
  doi:10.1016/0031-9163(64)91136-9.}{C1}
 
\xbibitem{Higgs:1964pj}{Higgs, P.W.,
  ``Broken Symmetries and the Masses of Gauge Bosons,''
  Phys.\ Rev.\ Lett.\  {\bf 13}, 508 (1964b).
  doi:10.1103/PhysRevLett.13.508.}{C1}

\xbibitem{Higgs:1966ev}{Higgs, P.W.,
  ``Spontaneous Symmetry Breakdown without Massless Bosons,''
  Phys.\ Rev.\  {\bf 145}, 1156 (1966).
  doi:10.1103/PhysRev.145.1156.}{C1}

\xbibitem{'tHooft:1972fi}{'t Hooft, G.\ and M.~J.~G.~Veltman, 
  ``Regularization and Renormalization of Gauge Fields,''
  Nucl.\ Phys.\ B {\bf 44}, 189 (1972).
  doi:10.1016/0550-3213(72)90279-9.}{C2}

\xbibitem{Jadach:1993hs}{Jadach, S., Z.~Was, R.~Decker and J.~H.~Kuhn,
  ``The tau decay library TAUOLA: Version 2.4,''
  Comput.\ Phys.\ Commun.\  {\bf 76}, 361 (1993).
  doi:10.1016/0010-4655(93)90061-G.}{C7}

\xbibitem{Kauer:2012hd}{Kauer, N.\ and G.~Passarino,
  ``Inadequacy of zero-width approximation for a light Higgs boson signal,''
  JHEP {\bf 1208}, 116 (2012)
  doi:10.1007/JHEP08(2012)116
  [arXiv:1206.4803 [hep-ph]].}{C4}

\xbibitem{Kersevan:2004yg}{Kersevan, B.P.\ and E.~Richter-Was,
  ``The Monte Carlo event generator AcerMC versions 2.0 to 3.8 with interfaces to PYTHIA 6.4, HERWIG 6.5 and ARIADNE 4.1,''
  Comput.\ Phys.\ Commun.\  {\bf 184}, 919 (2013)
  doi:10.1016/j.cpc.2012.10.032
  [hep-ph/0405247].}{C7}
  
\xbibitem{Kibble:1967sv} {Kibble, T.W.B.,
  ``Symmetry breaking in non-Abelian gauge theories,''
  Phys.\ Rev.\  {\bf 155}, 1554 (1967).
  doi:10.1103/PhysRev.155.1554.}{C1}

\xbibitem{Kunszt:1984ri}{Kunszt, Z.,
  ``Associated Production of Heavy Higgs Boson with Top Quarks,''
  Nucl.\ Phys.\ B {\bf 247}, 339 (1984).
  doi:10.1016/0550-3213(84)90553-4.}{C3}
  
\xbibitem{Lai:2010vv} {Lai, H.L., M.~Guzzi, J.~Huston, Z.~Li, P.~M.~Nadolsky, J.~Pumplin and C.-P.~Yuan,
  ``New parton distributions for collider physics,''
  Phys.\ Rev.\ D {\bf 82}, 074024 (2010)
  doi:10.1103/PhysRevD. 82.074024
  [arXiv:1007.2241 [hep-ph]].}{C7}
  
\xbibitem{Landau:1948kw}{Landau, L.D.,
``On the angular momentum of a system of two photons,'' 
Dokl.\ Akad.\ Nauk Ser.\ Fiz.\  {\bf 60}, no. 2, 207 (1948).
  doi:10.1016/B978-0-08-010586-4.50070-5.}{C6}

\xbibitem{Lee:1977eg}{Lee, B.W., C.~Quigg and H.~B.~Thacker,
  ``Weak Interactions at Very High-Energies: The Role of the Higgs Boson Mass,''
  Phys.\ Rev.\ D {\bf 16}, 1519 (1977).
  doi:10.1103/PhysRevD.16.1519.}{C2}

\xbibitem{LlewellynSmith:1973yud}{Llewellyn Smith, C.H.,
  ``High-Energy Behavior and Gauge Symmetry,''
  Phys.\ Lett.\ B {\bf 46}, 233 (1973).
  doi:10.1016/0370-2693(73)90692-8.}{C2}
    
\xbibitem{Mangano:2002ea}{Mangano, M.L., M.~Moretti, F.~Piccinini, R.~Pittau and A.~D.~Polosa,
  ``ALPGEN, a generator for hard multiparton processes in hadronic collisions,''
  JHEP {\bf 0307}, 001 (2003)
  doi:10.1088/1126-6708/2003/07/001
  [hep-ph/0206293].}{C7}

\xbibitem{Martin:2009iq}{Martin, A.D., W.~J.~Stirling, R.~S.~Thorne and G.~Watt,
  ``Parton distributions for the LHC,''
  Eur.\ Phys.\ J.\ C {\bf 63}, 189 (2009)
  doi:10.1140/epjc/s10052-009-1072-5
  [arXiv:0901.0002 [hep-ph]].}{C7}
  
\xbibitem{Melia:2011tj}{Melia, T., P.~Nason, R.~Rontsch and G.~Zanderighi,
  ``$W^+W^-$, $WZ$ and $ZZ$ production in the POWHEG BOX,''
  JHEP {\bf 1111}, 078 (2011)
  doi:10.1007/JHEP11(2011)078
  [arXiv:1107.5051 [hep-ph]].}{C5}

\xbibitem{Nadolsky:2008zw}{Nadolsky, P.M., H.~L.~Lai, Q.~H.~Cao, J.~Huston, J.~Pumplin, D.~Stump, W.~K.~Tung and C.-P.~Yuan,
  ``Implications of CTEQ global analysis for collider observables,''
  Phys.\ Rev.\ D {\bf 78}, 013004 (2008)
  doi:10.1103/PhysRevD.78.013004
  [arXiv:0802.0007 [hep-ph]].}{C7}
  
\xbibitem{Nason:2009ai}{Nason, P.\ and C.~Oleari,
  ``NLO Higgs boson production via vector-boson fusion matched with shower in POWHEG,''
  JHEP {\bf 1002}, 037 (2010)
  doi:10.1007/JHEP02(2010)037
  [arXiv:0911. 5299 [hep-ph]].}{C7}
  
\xbibitem{Passarino:2010qk}{Passarino, G., C.~Sturm and S.~Uccirati,
  ``Higgs Pseudo-Observables, Second Riemann Sheet and All That,''
  Nucl.\ Phys.\ B {\bf 834}, 77 (2010)
  doi:10.1016/j.nuclphysb.2010.03.013
  [arXiv: 1001.3360 [hep-ph]].}{C4}
  
\xbibitem{Rainwater:1997dg}{Rainwater, D.L.\ and D.~Zeppenfeld,
  ``Searching for $H\to\gamma\gamma$ in weak boson fusion at the LHC,''
  JHEP {\bf 9712}, 005 (1997)
  doi:10.1088/1126-6708/1997/12/005
  [hep-ph/9712271].}{C3}
 
\xbibitem{Rainwater:1998kj}{Rainwater, D.L., D.~Zeppenfeld and K.~Hagiwara,
  ``Searching for $H\to\tau^+\tau^-$ in weak boson fusion at the CERN LHC,''
  Phys.\ Rev.\ D {\bf 59}, 014037 (1998)
  doi:10.1103/PhysRevD.59.014037
  [hep-ph/9808468].}{C3}
  
\xbibitem{Rainwater:1999sd}{Rainwater, D.L.\ and D.~Zeppenfeld,
  ``Observing $H\to W^*W^* \to e^\pm \mu\mp$ + missing $p_T$  in weak boson fusion with dual forward jet tagging at the CERN LHC,''
  Phys.\ Rev.\ D {\bf 60}, 113004 (1999)
  Erratum: [Phys.\ Rev.\ D {\bf 61}, 099901 (2000)]
  doi:10.1103/PhysRevD.61.099901, 10.1103/PhysRevD.60.113004
  [hep-ph/9906218].}{C3}
  
\xbibitem{Ravindran:2003um}{Ravindran, V., J.~Smith and W.~L.~van Neerven,
  ``NNLO corrections to the total cross-section for Higgs boson production in hadron hadron collisions,''
  Nucl.\ Phys.\ B {\bf 665}, 325 (2003)
  doi:10.1016/S0550-3213(03)00457-7
  [hep-ph/0302135].}{C4}
  
\xbibitem{Salam:1968rm}{Salam, A.,
  ``Weak and Electromagnetic Interactions,''
  in {\it Elementary particle theory : relativistic groups and analyticity}, N.\ Svartholm (ed.), p.\ 367. 
  Almqvist \& Wiksell, 1968.
  Conf.\ Proc.\ C {\bf 680519}, 367 (1968).}{C1}

\xbibitem{Sherstnev:2007nd}{Sherstnev, A.\ and R.~S.~Thorne,
  ``Parton Distributions for LO Generators,''
  Eur.\ Phys.\ J.\ C {\bf 55}, 553 (2008)
  doi:10.1140/epjc/s10052-008-0610-x
  [arXiv:0711.2473 [hep-ph]].}{C7}
  
\xbibitem{Sjostrand:2006za}{Sjostrand, T., S.~Mrenna and P.~Z.~Skands,
  ``PYTHIA 6.4 Physics and Manual,''
  JHEP {\bf 0605}, 026 (2006)
  doi:10.1088/1126-6708/2006/05/026
  [hep-ph/0603175].}{C7}
  
\xbibitem{Sjostrand:2007gs}{Sjostrand, T., S.~Mrenna and P.~Z.~Skands,
  ``A Brief Introduction to PYTHIA 8.1,''
  Comput.\ Phys.\ Commun.\  {\bf 178}, 852 (2008)
  doi:10.1016/j.cpc.2008.01.036
  [arXiv:0710.3820 [hep-ph]].}{C7}
 
\xbibitem{Spira:1995rr}{Spira, M., A.~Djouadi, D.~Graudenz and P.~M.~Zerwas,
  ``Higgs boson production at the LHC,''
  Nucl.\ Phys.\ B {\bf 453}, 17 (1995)
  doi:10.1016/0550-3213(95)00379-7
  [hep-ph/9504378].}{C4}
  
\xbibitem{Stewart:2011cf}{Stewart, I.W.\ and F.~J.~Tackmann,
  ``Theory Uncertainties for Higgs and Other Searches Using Jet Bins,''
  Phys.\ Rev.\ D {\bf 85}, 034011 (2012)
  doi:10.1103/PhysRevD.85.034011
  [arXiv:1107. 2117 [hep-ph]].}{C4}

\xbibitem{Weinberg:1967tq}{Weinberg, S.,
  ``A Model of Leptons,''
  Phys.\ Rev.\ Lett.\  {\bf 19}, 1264 (1967).
  doi:10.1103/ PhysRevLett.19.1264.}{C2}
  
\xbibitem{Yang:1950rg}{Yang, C.N.,
  ``Selection Rules for the Dematerialization of a Particle Into Two Photons,''
  Phys.\ Rev.\  {\bf 77}, 242 (1950).
  doi:10.1103/PhysRev.77.242.}{C6}


\vfill\eject

\end{document}